# Quadratic shift-and-stack for Ground-Based Optical Detection of Faint Cislunar Objects

Qi Li[1], Yuhui, Zhao[1,2] [*], Chengxing Zhai[3], Yang Wang[1,2], Yi Han[1]

(1 Purple Mountain Observatory, Chinese Academy of Sciences, Nanjing 210023)
(2 School of Astronomy and Space Science, University of Science and Technology of China, Hefei 230026)
(3 Division of Emerging Interdisciplinary Areas, The Hong Kong University of Science and Technology, Clear Water Bay, Hong Kong)

**Abstract**

Detecting faint objects in cislunar space using ground-based optical telescopes is difficult because of their low brightness, strong lunar background, and complex, nonlinear apparent motion. Traditional shift-and-stack techniques based on linear motion assumption suffer signal trailing loss due to significant nonlinear motion during long integrations, thus producing degraded signal-to-noise ratio (SNR). In this paper, we first derive a theoretical criterion based on the point spread function to determine the maximum applicable integration time for a linear-motion stacking. We then propose a quadratic shift-and-stack (QSS) method to correct for the first order nonlinear motion, namely the angular acceleration of cislunar targets. Simulations of typical cislunar orbits verify this theoretical criterion and show that the QSS method significantly improves SNR from stacking and can enhance the detection limit by up to 1 stellar magnitude compared with the linear-motion stacking method. Furthermore, tests using observational data of the cislunar object Tiandu-1 confirm that while linear stacking degrades after a 29-minute integration due to trajectory curvature, the QSS method achieves continuous SNR improvement over a 46-minute integration, outperforming the peak SNR of the linear method by 31%.

[1] * Corresponding author: zhaoyuhui@pmo.ac.cn

# 1. Introduction

As lunar and deepspace exploration missions continue to expand, the cislunar region is becoming an increasingly critical strategic domain for space activities[1-3]. Many lunar missions and commercial payloads have increased the number of objects and the complexity of the cislunar environment[4,5]. Meanwhile, existing near-Earth observation and management systems are insufficient to effectively cover the cislunar region, resulting in increased risks of spacecraft collision, debris generation, and mission loss[6,7]. Establishing Cislunar Space Situational Awareness (CSSA) capability—timely monitoring of all space objects in the region—is fundamental to ensuring the safe and sustainable conduct of scientific, commercial, and crewed activities in cislunar space[1,8].

Ground-based optical sensors, due to their low cost, wide coverage, and ease of deployment, represent one of the most widely used observation methods in Space Situational Awareness (SSA)[9]. However, compared to near-Earth space, ground-based optical detection of cislunar objects faces several challenges. First, cislunar objects are much farther away and therefore appear much fainter than near-Earth objects[10]. In addition, these objects are often located near the Moon, where strong lunar background illumination further reduces the SNR. To address the challenge of low SNR, one might consider increasing exposure time to collect more photons from the object. However, traditional single-frame long-exposure methods are not suitable for cislunar detection because, like near-Earth asteroids (NEA), cislunar objects are both faint and fast-moving. Long exposures cause the object to trail across multiple pixels, which decreases the detection SNR[11]. Furthermore, under strong lunar background illumination, long exposures can result in image saturation, making object detection impossible and potentially risking detector damage.

The shift-and-stack technique offers an effective solution for detecting faint moving objects by avoiding trailing loss. The fundamental approach involves acquiring multiple short-exposure images and then computationally shifting and co-adding these frames according to the object's motion to synthesize an equivalent long-exposure image[12-14]. When frames are correctly aligned, the object signal accumulates constructively while noise grows only as the square root of the number of frames, resulting in SNR improvement proportional to the square root of the frame count[15]. This technique has been successfully applied to detecting faint NEAs and main-belt asteroids, demonstrating significant sensitivity improvements over conventional single-exposure methods[16,17].

The shift-and-stack method has undergone substantial development since its introduction in the early 1990s. Tyson et al.[18] and Cochran et al.[19] first applied the technique for trans-Neptunian object searches using Hubble Space Telescope data. Subsequent ground-based applications by Gladman et al.[12,20], Allen et al.[21], and Fraser et al.[22] demonstrated its effectiveness for detecting faint Kuiper Belt objects with large ground-based telescopes. For asteroid detection, Wang et al.[23] applied shift-and-stack methods to detect main-belt asteroids fainter than 21st magnitude using a near-Earth object survey telescope, while Heinze et al.[24] achieved detections down to R=23.4 magnitude using a 0.9-meter telescope. More recently, advances in camera technology and computing power have enabled "synthetic tracking"—a GPU-accelerated implementation of shift-and-stack that can search large velocity spaces in near real-time. Shao et al.[11] and Zhai et al.[15,16,25, 26] demonstrated synthetic tracking for detecting fast-moving near-Earth asteroids and reaching visual mag of 23 by integrating 1.5 hour Zwicky Transient Facility[27] deep drilling data, while Burdanov et al.[28] applied it to exoplanet survey data, reaching limiting magnitudes of V=23.8.

Despite these successes, existing shift-and-stack implementations predominantly rely on a linear motion model, assuming constant angular velocity over the observation duration. While this assumption is valid for objects at large geocentric distances that exhibit negligible trajectory curvature over typical observation windows, cislunar space presents a very different dynamical environment. Objects in this region are subject to strong, coupled gravitational perturbations from both the Earth and the Moon, resulting in highly nonlinear apparent motion on the focal plane even over timescales as short as 5 to 10 minutes. Under such conditions, the linear approximation breaks down. The uncorrected angular acceleration causes the target signal to drift across pixels during the stacking process, leading to trailing loss and imposing a hard limit on the effective integration time.

In this paper, we propose a quadratic shift-and-stack (QSS) method designed for detecting faint targets with variable acceleration. First, we theoretically derive a quantitative criterion based on the point spread function (PSF) to determine the maximum integration time for which the linear approximation remains valid. Second, we verify this theoretical limit through numerical simulations of typical cislunar orbits and evaluate the performance of the QSS method by quantifying the enhancement in the detection limit. Finally, we validate our approach using real observational data of the cislunar object Tiandu-1, acquired by the Multi-Application Survey Telescope Array (MASTA) at the Purple Mountain Observatory (PMO). The results demonstrate that QSS effectively compensates for acceleration, enabling continuous SNR improvement. Specifically, QSS achieves an SNR of 21.98 at 46 minutes, beyond the linear method's peak of 16.77 at 29 minutes.

The remainder of this paper is organized as follows: Section 2 details the theoretical limitations of the linear shift-and-stack method and introduces the proposed QSS method along with its applicability criterion. Section 3 presents extensive numerical simulations to evaluate the performance of QSS under various cislunar orbits. Section 4 validates the QSS using real observational data of the Tiandu-1 cislunar object. Finally, Section 5 concludes the paper.

## 2. Method

### 2.1 Linear Shift-and-Stack: Principle and Limitations

The shift-and-stack technique compensates for object motion by shifting consecutive frames before co-adding them. Assuming a sequence of $N$ frames is acquired, denoted as $I_n(x, y)$ where $n = 0, 1, \ldots, N - 1$ represents the frame index and $(x, y)$ are the pixel coordinates. If the object moves across the focal plane with a constant apparent velocity $(v_x, v_y)$, the linear shift-and-stack process aligns the frames by compensating for the displacement relative to a reference time $t_{\text{ref}}$. The synthesized image $S_{\text{linear}}$ is mathematically expressed as:

$$S_{\text{linear}}(x, y, \mathbf{v}) = \sum_{n=0}^{N-1} I_n(x + \Delta x_n, y + \Delta y_n) \tag{1}$$

where the shift vector $(\Delta x_n, \Delta y_n)$ is determined by the linear motion model[29,30]:

$$\Delta x_n = v_x(t_n - t_{\text{ref}}), \quad \Delta y_n = v_y(t_n - t_{\text{ref}}) \tag{2}$$

Here, $t_n$ is the timestamp of the $n$-th frame. To handle fractional pixel displacements during the shifting process, image resampling can be performed using bilinear interpolation before co-adding the frames. By searching through the velocity space $(v_x, v_y)$, the algorithm identifies the velocity vector that maximizes the target's intensity in the stacked image. This method assumes the target follows a linear trajectory with a constant apparent velocity throughout the observation.

However, this assumption does not always apply to objects in cislunar space. When significant angular acceleration is present, the target's position at time $t_n$ is more accurately modeled as:

$$\begin{aligned}
x_n(t) &\approx x_0 + v_x(t_n - t_{ref}) + \frac{1}{2}a_x(t_n - t_{\text{ref}})^2 \\
y_n(t) &\approx y_0 + v_y(t_n - t_{ref}) + \frac{1}{2}a_y(t_n - t_{\text{ref}})^2
\end{aligned} \tag{3}$$

where $(a_x, a_y)$ denotes the angular acceleration vector. When the linear shift-and-stack method Equation (1) is applied to such targets, the velocity compensation in Equation (2) only addresses the first-order displacement. The uncorrected residuals caused by the quadratic term spread the signal energy across multiple pixels instead of converging it into a compact PSF. This trailing affects the increase in SNR as integration time grows. Once the residual displacement exceeds the width of the system's PSF, adding more frames decreases the peak SNR. To overcome this limitation, a higher-order stacking model that explicitly accounts for angular acceleration is required.

**2.2 Quadratic Shift-and-Stack: Method and Applicability Criterion**

To address the trailing caused by angular acceleration, we propose a QSS that incorporates second-order terms into the trajectory compensation. The synthesized image $S_{\text{quad}}$ is constructed as:

$$S_{\text{quad}}(x, y, \mathbf{v}, \mathbf{a}) = \sum_{n=0}^{N-1} I_n(x + \Delta \tilde{x}_n, y + \Delta \tilde{y}_n) \tag{4}$$

The modified shift vector $(\Delta \tilde{x}_n, \Delta \tilde{y}_n)$ accounts for the influence of both velocity and acceleration:

$$\begin{aligned} \Delta \tilde{x}_n &= v_x(t_n - t_{\text{ref}}) + \frac{1}{2}a_x(t_n - t_{\text{ref}})^2 \\ \Delta \tilde{y}_n &= v_y(t_n - t_{\text{ref}}) + \frac{1}{2}a_y(t_n - t_{\text{ref}})^2 \end{aligned} \quad (5)$$

By introducing the quadratic compensation term $\frac{1}{2}a(t_n - t_{\text{ref}})^2$, this method maps the object signal from all frames back to the reference position with sub-pixel accuracy, eliminating residual trailing due to nonlinear motion and restoring the optimal detection SNR. In practice, the acceleration parameters $(a_x, a_y)$ can be obtained from ephemeris for a known target or through a parameter space search for unknown targets.

Although the QSS better matches the actual apparent motion of objects, the linear model remains a valid approximation for short exposures or targets with small angular accelerations. Therefore, it is necessary to theoretically derive the conditions under which the linear shift-and-stack model is no longer applicable. The following derivation is presented in one dimension for clarity; the extension to two-dimensional motion is addressed at the end of this section. The detailed derivation of the maximum residual displacement caused by angular acceleration is presented in Appendix A.

As derived in Appendix A, for a target moving with constant angular acceleration $a$ over a total integration time $T$, the maximum residual displacement $\delta_{\max}$ occurs at the observation boundaries and equals:

$$\delta_{\max} = \frac{1}{12}|a|T^2 \quad (6)$$

This result indicates that the alignment error caused by acceleration grows quadratically with integration time and is independent of the target's initial velocity. The decision to switch from linear to quadratic stacking depends on whether $\delta_{\max}$ degrades the peak SNR. As long as $\delta_{\max}$ remains within the core of the PSF, the photon energy continues to add coherently, and the SNR improves. Once $\delta_{\max}$ exceeds the PSF's width, the peak intensity saturates. We adopt the Full Width at Half Maximum (FWHM) of the system's PSF as the critical threshold. At this threshold, the trailing-induced sensitivity loss is approximately 12%[16], marking the onset of significant SNR degradation. Therefore, quadratic correction becomes necessary when:

$$\frac{1}{12}|a|T^2 > \text{FWHM} \quad (7)$$

Solving for $T$ defines the upper limit for the integration time $T_{\text{limit}}$:

$$T_{\text{limit}} = \sqrt{\frac{12 \cdot \text{FWHM}}{|a|}} \quad (8)$$

For observations where $T \leq T_{\text{limit}}$, the linear shift-and-stack method is applicable. For $T > T_{\text{limit}}$, or for objects with large angular accelerations, the linear approximation fails to confine the signal energy within the PSF and the QSS method must be used.

The above derivation was presented in one dimension for clarity. For two-dimensional apparent motion with angular acceleration components $(a_x, a_y)$, the residual displacement along each axis follows the same functional form as Equation (A7) in Appendix A. The maximum total residual displacement at the observation boundary is $\delta_{\max} = \frac{1}{12}|\boldsymbol{a}|T^2$, where $\boldsymbol{a} = \sqrt{a_x^2 + a_y^2}$ is the magnitude of the angular acceleration vector.

## 3. Simulation

In this section, we evaluate the performance of the proposed QSS method through numerical simulations. First, we analyze the apparent angular acceleration characteristics of typical cislunar orbits to understand the dynamical environment. Next, we simulate the shift-and-stack process to demonstrate the impact of angular acceleration and integration time on the linear stacking method, and we showcase the effectiveness of the quadratic correction. Finally, we systematically verify the theoretical criterion for the maximum effective integration time derived in Section 2.2, and further analyze the improvement in the detection limit provided by the quadratic stacking method.

### 3.1 Apparent Angular Acceleration Characteristics of Typical Cislunar Orbits

This section analyzes the apparent angular acceleration of objects operating in four typical cislunar orbits during ground-based observations. The ephemeris data originate from four operational spacecraft: DRO-A (Distant Retrograde Orbit, DRO), Tiandu-1 (Earth-Moon resonant orbit), Queqiao-1 (Earth-Moon L2 halo orbit), and Queqiao-2 (Lunar orbit). The data consists of Right Ascension and Declination coordinates from June 1 to June 30, 2025, with a temporal resolution of one second. Using the finite difference method, we calculate the apparent angular velocity ("/min) and apparent angular acceleration ("/min$^2$) at each epoch. Figure 1 illustrates the variation of angular acceleration for these four objects over a two-day period from June 19 to June 20, 2025, while Table 1 provides the statistics for the entire month of June 2025.

The upper panel of Figure 1 shows the angular acceleration for the resonant and lunar orbit objects. Both exhibit relatively large angular accelerations (reaching over 1"/min$^2$). The lower panel displays the DRO and halo orbit objects, which have much smaller angular accelerations (generally below 0.1"/min$^2$). As seen in the figure, the angular acceleration of the lunar object shows clear periodicity and dramatic magnitude changes. As detailed in Table 1, its minimum angular acceleration in June is 0.0013"/min$^2$, while the maximum reaches 2.3363"/min$^2$, with an average of 0.1366"/min$^2$. The resonant orbit object also exhibits a wide range of angular acceleration, from a minimum of 0.0062"/min$^2$ to a maximum of 1.3542"/min$^2$, averaging 0.1503"/min$^2$. The DRO and halo objects also show distinct periodicity, but their overall numerical variations are much smaller. For the DRO object, the minimum angular acceleration is 0.0002"/min$^2$, the maximum is 0.3792"/min$^2$, and the average is 0.0402"/min$^2$. For the halo orbit object, the minimum is 0.0004"/min$^2$ and the maximum is 0.2717"/min$^2$.

These statistics reveal significant differences in apparent angular acceleration among different types of cislunar orbits. Furthermore, the apparent acceleration of an object in the same orbit varies greatly depending on its position, reflecting the complexity of apparent motion in cislunar space, fundamentally driven by the complex Earth-Moon three-body dynamical environment. Analyzing these kinematic characteristics can guide object detection strategies. For instance, Table 1 shows that the apparent velocity of the DRO object ranges from 7.1 to 71.33"/min. This prior knowledge can restrict the velocity search space during shift-and-stack detection of uncooperative DRO objects. Similarly, the statistical distribution of angular acceleration provides valuable prior information for deploying the QSS method proposed in this paper. Adding acceleration parameters increases the search space by two dimensions, significantly increasing computational load. Utilizing prior information can effectively constrain the parameter search range, thereby reducing computational cost.

This section has quantified the apparent angular acceleration of four types of cislunar objects. The subsequent simulations and real-data processing will visually demonstrate the impact of this angular acceleration on the shift-and-stack process.

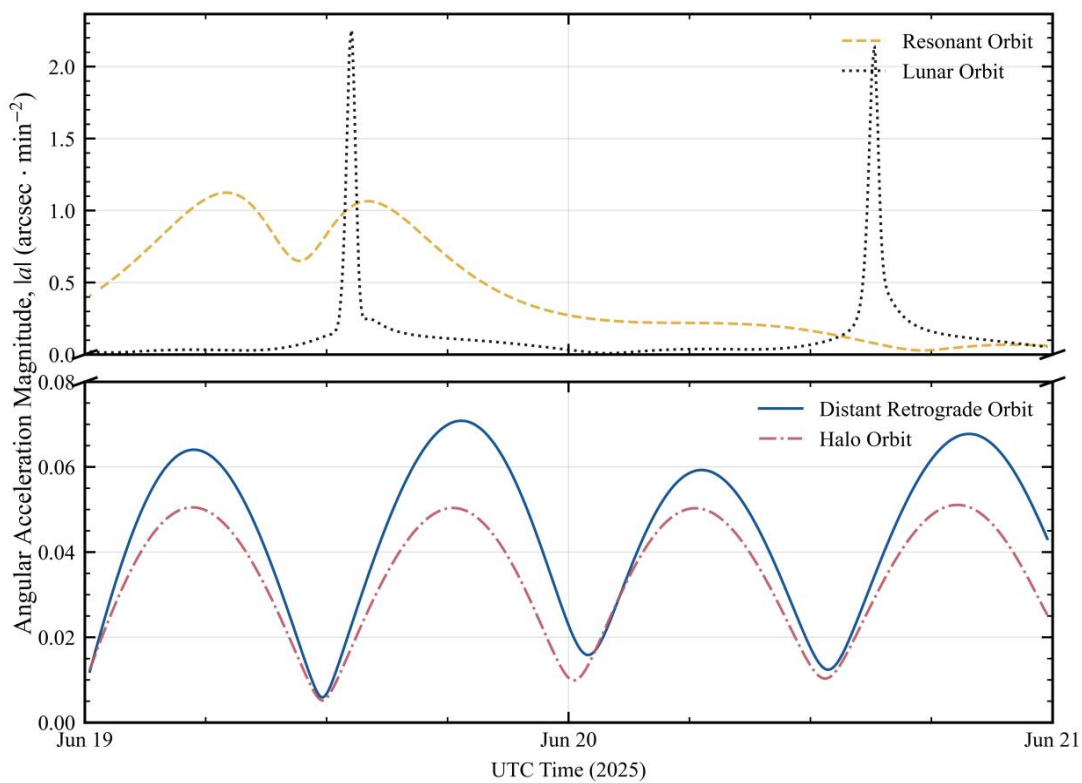

**Figure 1**. Variation of apparent angular acceleration for four types of cislunar orbit objects during ground-based observations from June 19 to June 20, 2025.

**Table 1**. Statistics of apparent angular velocity("/min) and angular acceleration("/min$^2$) for four types of cislunar orbit objects during ground-based observations from June 1 to June 30, 2025.

| Orbit | $v_{min}$ | $v_{max}$ | $v_{mean}$ | $v_{std}$ | $a_{min}$ | $a_{max}$ | $a_{mean}$ | $a_{std}$ |
|---|---|---|---|---|---|---|---|---|
| Resonant orbit | 6.62 | 836.22 | 97.89 | 147.99 | 0.0062 | 1.3542 | 0.1503 | 0.2442 |
| Lunar orbit | 1.59 | 108.83 | 38.84 | 13.66 | 0.0013 | 2.3363 | 0.1366 | 0.2762 |
| DRO | 7.10 | 71.33 | 33.71 | 14.42 | 0.0002 | 0.3792 | 0.0402 | 0.0146 |
| Halo orbit | 16.24 | 54.58 | 34.03 | 8.85 | 0.0004 | 0.2717 | 0.0348 | 0.0124 |

## 3.2 Analysis of Shift-and-Stack Performance under Angular Acceleration using Simulation

In this section, we use image stacking simulations (the simulation method and parameter configurations are detailed in Appendix B) to analyze the impact of apparent angular acceleration and integration time on linear shift-and-stack. We demonstrate the effectiveness of QSS and verify the theoretical boundary for linear stacking validity given by Equation (8).

### 3.2.1 Impact of Angular Acceleration on Linear Shift-and-Stack

As shown in Section 3.1, the apparent angular acceleration of cislunar objects varies over a wide range (from 0.002 to 2."/min$^2$). This subsection demonstrates the impact of angular acceleration on linear stacking through two comparative simulation cases.

The first case (Case 1) represents a scenario with small apparent angular acceleration. We selected ephemeris data for Queqiao-2 (Lunar orbit) over a 10-minute period from 10:20:00 to 10:30:00 UTC on June 8, 2025. During this period, the object's apparent angular velocity was 36.63"/min with a position angle of 103.0º (measured East of celestial North), and its apparent angular acceleration was 0.053"/min$^2$ with a position angle of 314.1º. The object's trajectory on the image focal plane over these 10 minutes is shown by the solid pink line in Figure 2(a). The coordinates in the trajectory plot are in pixels (pixel scale is 1.67"/pixel), where the positive X direction (right side of the image) corresponds to celestial East, and the positive Y direction (bottom side of the image) corresponds to celestial North. The celestial coordinates from the ephemeris (one data point per second) were converted to image coordinates via gnomonic projection and scaled by the pixel scale. The start and end points of the trajectory are marked with a dot and a cross, respectively. As seen in Figure 2(a), with an angular acceleration of 0.053"/min$^2$, the trajectory over 10 minutes is nearly a straight line. The solid blue line and the dashed green line represent the best-fit linear and quadratic approximations to this trajectory, respectively. Both linear and quadratic fits match the actual trajectory very well.

Next, we performed image shift-and-stack simulations. The 10-minute ephemeris provides 601 position data points (one per second). We simulated observations with a 1-second exposure time by generating 601 simulated image frames. Each frame contains background noise and the object PSF (which also includes photon noise following a Poisson distribution). The background noise includes sky background noise (1000 e$^-$/pixel/s, Poisson distribution), dark current noise (0.05 e$^-$/pixel/s, Poisson distribution), and readout noise (2 e$^-$/pixel/frame, Gaussian distribution). The object is modeled as a Gaussian PSF with a peak intensity of 10 e$^-$ (simulated gain is 1, so equivalent to 10 ADU) and a FWHM of 3". Given the pixel scale of 1.67"/pixel, the object's peak SNR in a single frame is 0.36 (invisible). We applied both the linear shift-and-stack and the QSS methods to these 601 frames. The results are shown in Figures 2(b) and 2(c). The object becomes visible after both linear and quadratic stacking. The SNR of the quadratic stack (8.79) is slightly higher than that of the linear stack (8.53). The figures also display noise-free versions of the stacked PSF shapes, showing that both methods concentrate the object into a single point source. Case 1 demonstrates that for a small angular acceleration of 0.053"/min$^2$ and a 10-minute integration time, linear shift-and-stack can effectively align the object and improve the SNR.

The second case (Case 2) represents a scenario with large apparent angular acceleration. We selected Queqiao-2 ephemeris data for the 10-minute period from 17:10:00 to 17:20:00 UTC on June 8, 2025. The object's apparent angular velocity was 35.80"/min with a position angle of 301.2º, and its angular acceleration was 1.659"/min$^2$ with a position angle of 224.9º. As shown in the trajectory plot in Figure 2(d), the high angular acceleration causes the trajectory to curve significantly. Here, the linear fit fails to match the trajectory, while the quadratic fit aligns well. We performed the same image stacking simulation for this period (using identical background and object parameters). The results in Figures 2(e) and 2(f) reveal that the object remains undetectable after linear stacking with a SNR of 2.00. The noise-free PSF plot shows that the object energy is not aligned but instead forms a distinct trail. Conversely, the quadratic stack concentrates the object into a point source, achieving an SNR of 8.45 and making the object clearly visible.

These two cases demonstrate that when apparent angular acceleration is small, linear shift-and-stack can adequately align the object across frames to improve SNR. However, when the angular acceleration is large, linear stacking fails to align the object, resulting in significant trailing loss, which prevents SNR improvement. In contrast, QSS effectively accounts for acceleration to align frames and achieves SNR enhancement.

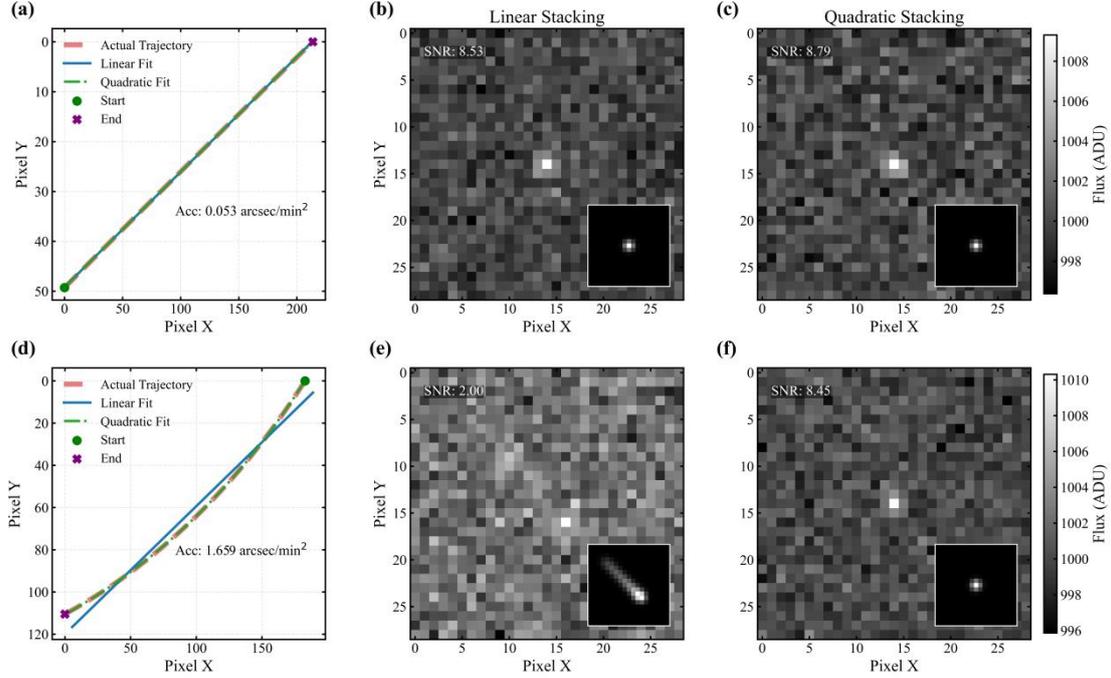

**Figure 2**. Performance of linear and quadratic shift-and-stack under small and large apparent angular accelerations: (a-c) object trajectory, linear stacked image, and QSS image under small angular acceleration (0.053″/min$^2$), simulated using Queqiao-2 ephemeris from 10:20:00 to 10:30:00 UTC on June 8, 2025; (d-f) object trajectory, linear stacked image, and quadratic stacked image under large angular acceleration (1.659″/min$^2$), simulated using Queqiao-2 ephemeris from 17:10:00 to 17:20:00 UTC on June 8, 2025.

### 3.2.2 Impact of Integration Time on Linear Shift-and-Stack

We now discuss the impact of integration time on linear shift-and-stack. In Case 1 (Section 3.2.1), where the angular acceleration was 0.053″/min$^2$, linear stacking was effective for a 10-minute integration. We extended the simulation using 50 minutes of ephemeris data for the same object (Queqiao-2, June 8, 2025, 10:20:00 to 11:10:00). We performed stacking simulations starting with a 1-minute integration time and increasing it up to 50 minutes in 1-minute increments. The SNR of the object after linear and quadratic stacking was recorded for each duration.

Figure 3 shows the SNR as a function of the integration time (minutes). The dashed black line represents the theoretical ideal SNR, assuming perfect alignment, calculated by multiplying the single-frame SNR (0.36 for a 1-second exposure) by the square root of the integration time. The blue line shows the SNR with linear shift-and-stack. For short integration times (e.g., the first 10 minutes), the linear stacking SNR increases with time, closely following the theoretical curve. However, as the integration time increases further, the linear stacking SNR reaches a maximum and then begins to decrease. For this specific case, linear stacking achieves a maximum SNR of 11.37 at 25 minutes. This empirical maximum is consistent with the theoretical integration limit of 26.06 minutes calculated using Equation (8), with a relative error of 4.1%. At this point, the PSF image shows slight trailing. By 50 minutes, the SNR drops to 5.66, and the object exhibits severe trailing. This phenomenon is physically intuitive: with a small angular acceleration, the object's trajectory is nearly linear over short integration times, resulting in small trailing errors for linear stacking (Equation (6)). However, over longer integration times, the trajectory's curvature becomes significant. Linear stacking suffers large trailing loss, causing the SNR to drop.

The red line in Figure 3 represents the SNR variation of QSS, which closely follows the theoretical curve throughout. At 50 minutes, the SNR reaches 19.49, which is only 0.16 lower than the theoretical value (19.65)—this slight degradation is primarily due to unmodeled higher-order nonlinear motions (such as jerk)—and 71.41% higher than the peak linear stacking SNR (11.37). The PSF images confirm that quadratic stacking properly compensates for acceleration, keeping the object focused as a point source.

This case demonstrates that the impact of apparent angular acceleration on linear shift-and-stack depends on both the magnitude of the acceleration and the integration time. For a given angular acceleration, a longer integration time leads to larger trailing errors, eventually causing linear stacking to fail beyond a critical integration limit. Only by incorporating acceleration compensation during the shifting process can the integration time be continuously extended to improve object SNR.

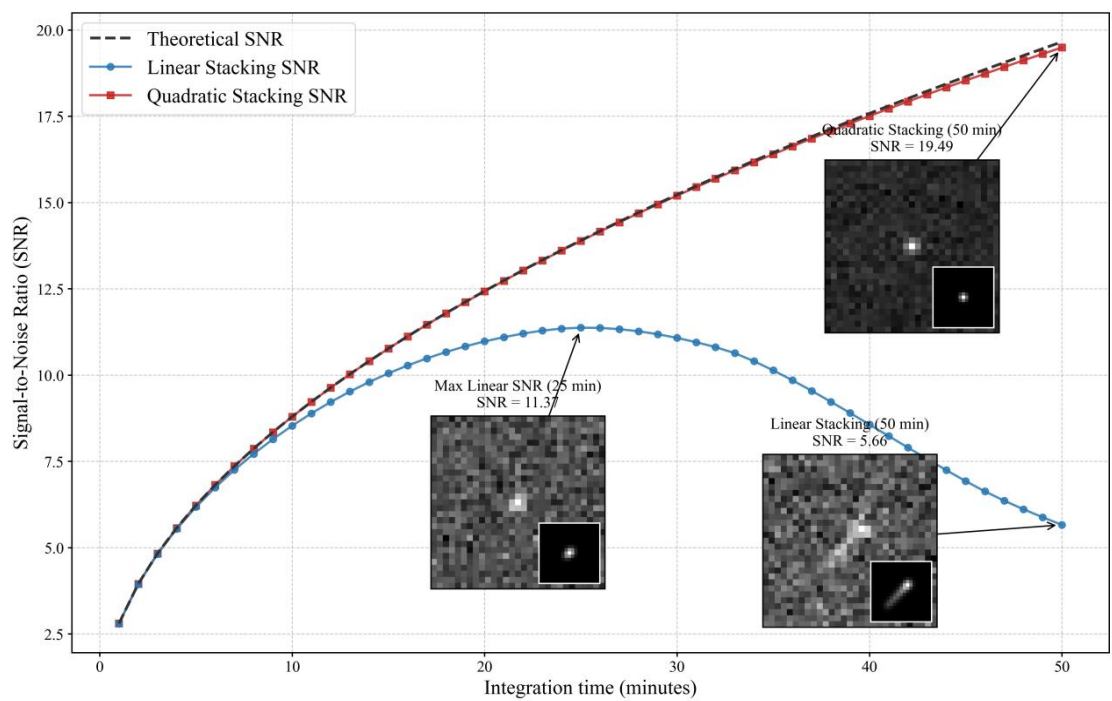

**Figure 3**.Simulated SNR variation of linear and quadratic stacking with integration time, using 50 minutes of Queqiao-2 ephemeris data from 10:20:00 to 11:10:00 UTC on June 8, 2025 (object apparent angular acceleration is 0.053"/min$^2$).

### 3.2.3 Theoretical Verification of the Impact of Angular Acceleration and Integration Time

The previous subsections discussed the individual impacts of angular acceleration and integration time using the lunar orbit ephemeris. In this section, we systematically analyze the combined effects of angular acceleration and integration time on both linear and quadratic stacking via extensive simulations. The procedure is as follows: We varied the integration time from 0.5 minutes to 60 minutes in 0.5-minute steps. For each integration time, we randomly selected 4,000 data sequences from the four orbital ephemerides (DRO, halo, resonant, and lunar, each containing data for the entire month of June 2025). We performed linear and quadratic stacking simulations for these 4,000 cases at each time step, recording the resulting SNRs. We then mapped these results against the apparent angular acceleration of each case to create the heatmaps shown in Figure 4 (comprising 480,000 simulation cases in total). The angular acceleration axis is divided into 60 bins; if a bin contains multiple results, the average SNR is plotted. The acceleration range is 0-1.35"/min$^2$ (data below 1.35 includes all four orbit types, while data above 1.35 only includes lunar orbit).

Figure 4(a) shows the SNR distribution for linear shift-and-stack across different integration times and angular accelerations. The overall distribution indicates that, regardless of the angular acceleration, the linear stacking SNR initially increases with integration time and then decreases. The larger the angular acceleration, the narrower the "yellow-green" region, indicating poorer linear stacking performance. Notably, for angular accelerations greater than 0.1"/min$^2$, a large dark-blue region emerges, showing that linear stacking almost completely fails at higher accelerations. The dashed white line in Figure 4(a) represents the theoretical integration limit described by Equation (8); beyond this time, the linear stacking SNR begins to drop. The red crosses mark the maximum SNR point for each row (each acceleration value). These empirical maximums align almost perfectly with the theoretical curve of Equation (8) (the 3.23% error is primarily due to the discrete 0.5-minute simulation time steps, whereas the theoretical curve is continuous). This confirms the correctness of the integration limit theory described by Equation (8). The practical value of this equation is that, given an object's apparent angular acceleration, one can calculate the maximum effective integration time for linear shift-and-stack.

Figure 4(b) shows the SNR distribution for QSS. The lower-right region is bright yellow (SNR > 20), indicating that for small accelerations, the quadratic stacking SNR continues to increase significantly with integration time. Most of the map is yellow-green, with a dark blue region only in the upper right corner. This demonstrates that quadratic stacking is suitable for long-integration detection of objects in all four orbit types. The dark blue regions indicate extreme cases where objects exhibit higher-order motion dynamics, such as jerk (changing acceleration), rendering even quadratic stacking insufficient. Furthermore, in Figure 4(b), at higher angular accelerations, while the overall trend still shows SNR increasing and then decreasing, the detailed distribution is somewhat noisy. This is because higher-order terms (such as jerk) are not accounted for in QSS. This leads to "anomalies" where, for instance, two cases with the same angular acceleration may exhibit different quadratic-stacking SNRs if their higher-order motion terms differ.

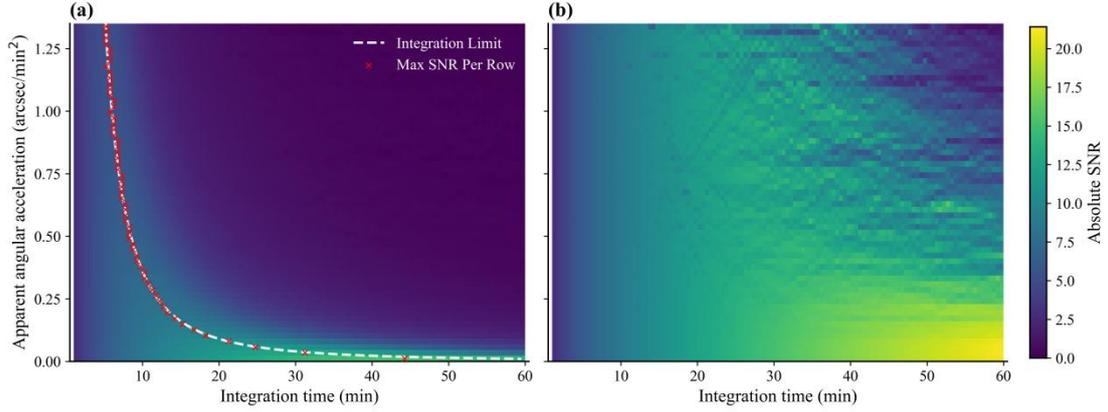

**Figure 4.** SNR distribution of linear (a) and quadratic shift-and-stack (b) under varying angular accelerations and integration times.

To further evaluate the practical improvement of quadratic stacking over linear stacking for cislunar object detection, we compared the stacked SNR of each point in Figure 4 to its theoretical SNR (the SNR under perfect alignment), resulting in the distributions shown in Figure 5. We define the ratio of the stacked SNR to the theoretical ideal SNR as the "Stacking Efficiency" (SE). SE ranges from 0 to 1; a higher value indicates more effective stacking (i.e., more accurate object alignment). Figure 5(a) shows that linear shift-and-stack is only highly efficient for short integration times, and this effective window shrinks rapidly as angular acceleration increases. We regard SE > 0.8 as indicating highly efficient stacking, the contour line for linear SE = 0.8 in Figure 5(a) nearly overlaps with the theoretical integration limit curve. This means that when linear stacking reaches its maximum SNR, that value is approximately 80% of the theoretical ideal SNR for that integration time. Compared to linear stacking, the effective region for quadratic stacking is significantly expanded, with the vast majority of the lower-left area appearing yellow.

Quantitative statistics on stacking efficiency are presented in Table 2, which provides breakdowns by orbit type as well as overall totals. Figure 5 represents the aggregate distribution across all four orbit types. Overall, linear stacking achieves an SE > 0.8 in only 17.59% of cases, whereas quadratic stacking achieves this in 64.50% of cases—an absolute increase of 46.91%. This demonstrates that QSS is much better suited for detecting objects with complex apparent motion in cislunar space. The mean SE across all four orbits is 0.2873 for linear stacking and 0.7480 for quadratic stacking, representing a 160.3% improvement.

Looking at the statistics for different types of orbits, linear stacking has relatively higher efficiency for DRO and halo orbits (mean SE of 0.6078 and 0.5251, respectively) and lower efficiency for resonant and lunar orbits (0.2699 and 0.2644). This is because the latter two orbits exhibit higher angular accelerations (as shown in Table 1). For quadratic stacking, the efficiency is higher for resonant and DRO orbits (0.8637 and 0.8065) and relatively lower for lunar and halo orbits (0.6258 and 0.7022). We hypothesize that this is due to more prominent higher-order apparent motion dynamics (such as jerk) for objects in lunar and halo orbits.

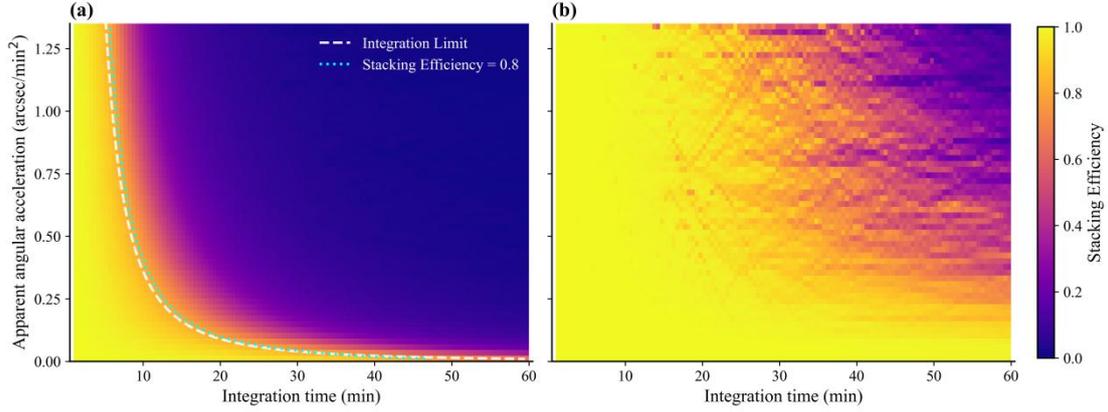

**Figure 5.** Distribution of the ratio between stacked SNR and theoretical optimal SNR (Stacking Efficiency) for linear and quadratic shift-and-stack.

**Table 2.** Statistics of linear and quadratic stacking efficiency for different types of cislunar orbit objects.

| Orbit | Linear SE > 0.8 | Mean Linear SE | Quadratic SE > 0.8 | Mean Quadratic SE |
|---|---|---|---|---|
| Resonant orbit | 15.86% | 0.2699 | 77.48% | 0.8637 |
| Lunar orbit | 15.92% | 0.2644 | 50.89% | 0.6258 |
| DRO | 44.73% | 0.6078 | 73.83% | 0.8065 |
| Halo orbit | 37.74% | 0.5251 | 54.88% | 0.7022 |
| Total | 17.59% | 0.2873 | 64.50% | 0.7480 |

### 3.3 Enhancement of Detection Limit

Previous analyses have shown that under the influence of apparent angular acceleration, linear shift-and-stack fails to improve the target's SNR once the integration time reaches a critical limit, whereas QSS can further enhance the SNR. An increase in SNR directly translates to the ability to detect fainter objects. This section evaluates the extent to which quadratic stacking improves the detection limit.

The analysis method is as follows: We selected segments of ephemeris data from the resonant orbit with specific angular accelerations $a$, each containing 60 minutes of data. Both linear and quadratic shift-and-stack simulations were then performed with progressively increasing integration times. Unlike previous simulations where the peak flux of the target's PSF in a single frame was fixed, in this simulation, we dynamically adjusted the target's peak flux for each stacking case so that the final stacked SNR exactly reached a detection threshold of 5. The total flux within a circular aperture of radius equal to 1 FWHM was then recorded, denoted as $Flux_{\mathrm{lin}}$ and $Flux_{\mathrm{quad}}$. The stellar magnitude difference (gain) between quadratic and linear stacking was calculated using the following equations:

$$\Delta m = m_{\text{quad}} - m_{\text{lin}} = 2.5 \times \log_{10} \left(\frac{Flux_{\text{lin}}}{Flux_{\text{quad}}}\right), T \leq T_{\text{limit}}$$
$$\Delta m = m_{\text{quad}} - m_{\text{lin}}^* = 2.5 \times \log_{10} \left(\frac{Flux_{\text{lin}}^*}{Flux_{\text{quad}}}\right), T > T_{\text{limit}} \tag{9}$$

Here, $m_{\text{quad}}$ and $m_{\text{lin}}$ represent the instrumental magnitudes of the target in the quadratic and linear stacked images, respectively, and $\Delta m$ is the stellar magnitude gain of quadratic stacking relative to linear stacking. Note that because the SNR of linear stacking stops increasing after the critical integration time $T_{\text{limit}}$, the optimal detection limit for linear stacking, $m_{\text{lin}}^*$, is achieved at $T_{\text{limit}}$. Therefore, for $T > T_{\text{limit}}$, the magnitude gain is calculated relative to $m_{\text{lin}}^*$. We selected data segments with angular accelerations of 0.02, 0.05, 0.15, 0.50, and 1.00 "/min$^2$ to perform these calculations. The relationship between magnitude gain and integration time is illustrated in Figure 6, and the corresponding values are listed in Table 3.

In Figure 6, the square markers on each curve indicate the $T_{\text{limit}}$ for that specific angular acceleration. For angular accelerations of 0.02, 0.05, and 0.15 "/min$^2$, the magnitude gain of quadratic stacking increases continuously with integration time. The optimal detection limits for linear stacking are reached at 42.4, 26.8, and 15.5 minutes, respectively. After these points, linear stacking becomes ineffective, while quadratic stacking continues to improve the detection limit, reaching magnitude gains of 0.425, 0.650, and 0.933 at 60 minutes. For larger angular accelerations of 0.50 and 1.00 "/min$^2$, the magnitude gain initially increases but then decreases. This indicates that after a certain integration time, quadratic stacking also begins to fail, primarily because higher-order apparent motion dynamics (such as jerk) introduce residual trailing errors. However, compared to scenarios with smaller accelerations, the magnitude gain achieved by quadratic stacking over short duration is much more pronounced for larger accelerations, reaching maximum gains of 0.692 and 1.064 at 33 and 34 minutes, respectively.

These results demonstrate that QSS effectively enhances the detection limit. For small angular accelerations, the magnitude gain increases steadily with longer integration time; for large angular accelerations, quadratic stacking provides a rapid and significant magnitude gain over shorter periods. Overall, the quadratic method can extend the detection limit by up to 1 stellar magnitude or more.

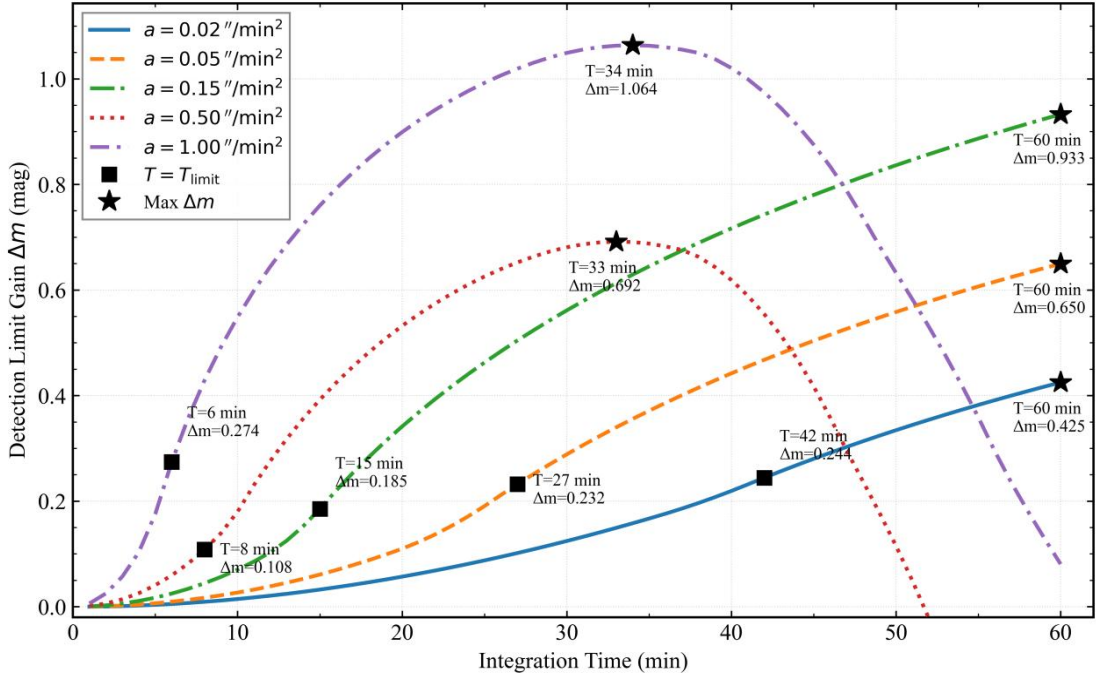

**Figure 6**. Stellar magnitude gain of QSS relative to the linear method as a function of integration time under different angular accelerations.

Table 3. Stellar magnitude gain of QSS relative to the linear method.

| Angular Accel ("/min²) | $T_{\text{limit}}$ (min) | 10min | 20min | 30min | 40min | 50min | 60min | Max $\Delta m$ |
|---|---|---|---|---|---|---|---|---|
| 0.02 | 42.4 | 0.015 | 0.057 | 0.124 | 0.219 | 0.334 | 0.425 | 0.425 |
| 0.05 | 26.8 | 0.027 | 0.110 | 0.289 | 0.442 | 0.558 | 0.650 | 0.650 |
| 0.15 | 15.5 | 0.070 | 0.342 | 0.561 | 0.717 | 0.836 | 0.933 | 0.933 |
| 0.50 | 8.5 | 0.180 | 0.532 | 0.681 | 0.619 | 0.115 | -0.521 | 0.692 |
| 1.00 | 6.0 | 0.549 | 0.899 | 1.049 | 1.021 | 0.633 | 0.081 | 1.064 |

## 4. Application

### 4.1 Data Acquisition and Preprocessing

The observations were conducted using the MASTA telescope at the Purple Mountain Observatory. We observed the cislunar object Tiandu-1 on 18 December 2025 from 14:40:00 to 15:26:39 UTC. The telescope operated in sidereal tracking mode, acquiring a total of 800 frames with an exposure time of 3.0 seconds per frame. The readout time between consecutive frames was 0.5 s.

At the beginning of the observation window (14:40:00 UTC), the object was located at RA=30.129º and Dec = 6.920º. The telescope pointing was at an azimuth of 225.829º and an altitude of 64.625º. The object exhibited an apparent angular velocity of approximately 47.5"/min with a position angle (PA) of 41.5º. Due to the large field of view (FOV) of MASTA (3.75º×3.75º), the object remained within the FOV throughout the 46-minute session. Crucially, the object exhibited an apparent angular acceleration of 0.044"/min², PA=93.59º, making it an ideal candidate for testing the robustness of the nonlinear stacking algorithm.

The raw observational data were processed using a custom pipeline developed in Python, incorporating standard astronomical libraries such as *ccdproc*, *astropy*, and *sep*. The data reduction procedure consisted of the following steps:

First, standard instrumental calibration was applied to the raw frames. This involved bias subtraction, dark current subtraction (scaled by exposure time) to eliminate thermal noise, and flat-field division to correct for optical vignetting and non-uniform pixel response.

Subsequently, specific algorithms were implemented to address artifacts inherent to the CMOS detector, such as sensor gaps and bad columns. These anomalies were identified via row and column median statistics and corrected using interpolation from adjacent pixels. Simultaneously, hot pixels were identified and masked using a sigma-clipping algorithm with a threshold of $5\sigma$ and then interpolated using a local median filter.

Following these corrections, the sky background was estimated using the *sep* library (a Python implementation of Source Extractor) with a background mesh size of 64×64 pixels and subsequently subtracted from the original images. The images were then normalized, with pixel values rescaled to the [0,1] range. Finally, the pipeline computed the astrometric solution (WCS) and performed image alignment.

**4.2 Performance Comparison of Stacking Methods**

Following image preprocessing, an analysis was performed using both linear and quadratic shift-and-stack methods on the image sequence.

First, frames 791–800 were excluded from the dataset to avoid photometric contamination caused by the object blending with a background star. We then conducted stacking tests using cumulative subsets of the sequence: the first 1 min (frames 1 – 20), the first 2 min (frames 1 – 40), and so on, up to the first 46 min (frames 1 – 790).

The motion parameters (velocity and acceleration) required for the stacking algorithms were derived from fitting the object's ephemeris. For each stacked image, the SNR of the object was measured using a circular photometric aperture with a radius equal to 1 FWHM (3.0").

Figure 7 illustrates the results of linear and quadratic shift-and-stack with varying numbers of frames. It can be observed that for linear shift-and-stack, when 400 frames are stacked (spanning a total duration of 23 minutes, including a 3-second exposure per frame and a 0.5-second readout time between frames), the object appears as a point source with a SNR of 16.64. The SNR reaches its maximum value of 16.77 when the number of stacked frames increases to 500 (29 minutes). Beyond this point, the PSF of the object begins to smear (or diverge), leading to a decrease in SNR. Substituting the FWHM of 3" and the angular acceleration of 0.044"/min$^2$ into Equation (8) yields a theoretical integration limit $T_{\text{limit}} = \sqrt{\frac{12 \times 3}{0.044}} = 28.6$ minutes, which is approximately 29 minutes. This empirical result corroborates the correctness of the theoretical Equation (8).

In contrast, for quadratic shift-and-stack, the SNR consistently increases with the number of stacked frames. At 790 frames, the SNR reaches 21.98, representing an improvement of 31% compared to the peak SNR of 16.77 obtained from linear shift-and-stack. These results demonstrate that while linear shift-and-stack becomes ineffective for nonlinear moving objects over long integration times, quadratic shift-and-stack continues to work.

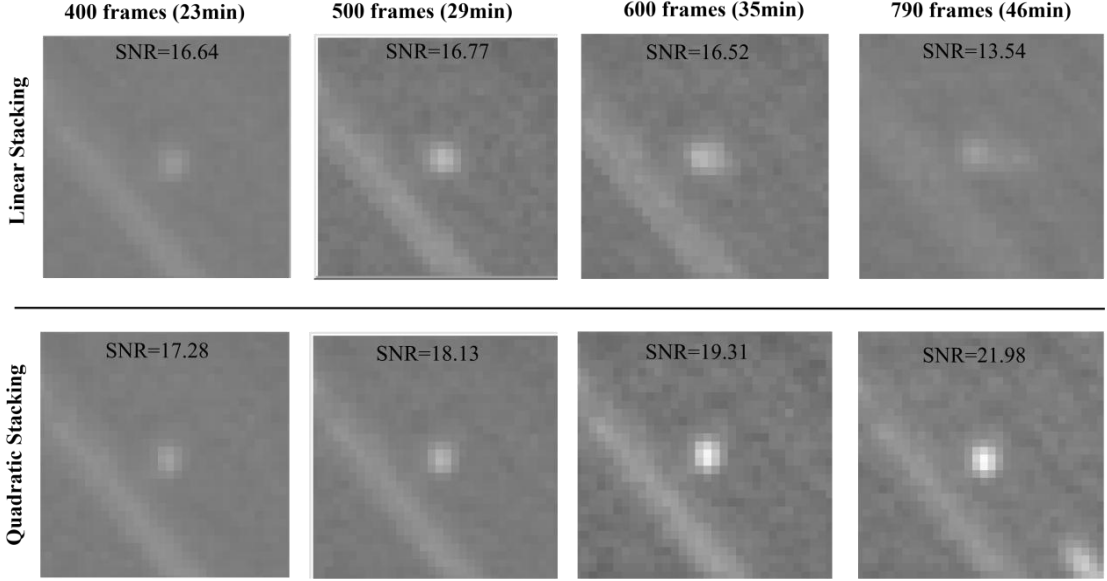

**Figure 7**. Comparison of linear and quadratic shift-and-stack results with different numbers of stacked frames for Tiandu-1 (2025-12-18 14:40:00 to 15:26:39). The theoretical integration limit for linear stacking is $T_{\text{limit}} = 28.6$ minutes (corresponding to approximately 500 frames).

## 5. Conclusions

Detecting faint cislunar objects with ground-based optical telescopes is challenging because of their low brightness, strong lunar background, and nonlinear apparent motion, which limits the performance of conventional stacking methods. In this paper, we proposed a quadratic shift-and-stack (QSS) method to improve the detection of targets with significant angular acceleration in cislunar space. To quantify the limitation of the conventional linear shift-and-stack method, we derived a PSF-based theoretical criterion, $T_{limit}$, for determining its maximum effective integration time. Simulations of representative cislunar orbits validated this criterion and showed that the QSS method can better compensate for nonlinear apparent motion during long integrations. As a result, QSS provides higher SNR gains than the linear method and can improve the detection limit by up to 1 stellar magnitude, depending on the target's angular acceleration and integration time.

The method was further evaluated using real observations of the cislunar object Tiandu-1. The results showed that the SNR of linear stacking began to degrade after about 29 minutes of integration because of uncorrected trajectory curvature, whereas the QSS method maintained signal coherence and continued to improve the SNR over a 46-minute integration. Under these conditions, QSS achieved a 31% higher peak SNR than the linear method. These results demonstrate that QSS is an effective approach for the optical detection of faint cislunar targets with noticeable nonlinear motion.

**Acknowledgments**

This work is supported by the Natural Science Foundation of Jiangsu Province (Youth Fund Project) under Grant No. BK20251704.

**Appendix A:**

The following derivation is presented in one dimension for clarity; the extension to two-dimensional motion is addressed at the end of Section 2.2.

Consider $N$ frames acquired at uniformly spaced times $t_n = n\Delta t$ ($n = 0, 1, \ldots, N-1$), with total observation duration $T = (N-1)\Delta t$. A target moving with constant acceleration $a$ has its position at $t_n$ given by:

$$x(t) = x_0 + v_0 t + \frac{1}{2} a t^2 \tag{A1}$$

where $x_0$ is the initial position and $v_0$ is the initial angular velocity. When using the linear shift-and-stack algorithm, it searches for an optimal constant matching velocity $v_{\text{opt}}$ to maximize the stacked peak energy. This is equivalent to minimizing the residual between the target's true position $x(t)$ and the linearly compensated position $x_{\text{linear}}(t)$ over the integration time $[0, T]$ [15,25]. By minimizing the mean-square positional residual $\text{var} = \frac{1}{T} \int_0^T [x(t) - x_{\text{linear}}(t)]^2$ over $[0, T]$, the optimal matching velocity is found to be the instantaneous velocity at the midpoint:

$$v_{\text{opt}} = v_0 + a\left(\frac{T}{2}\right) = v_0 + \frac{1}{2} a T \tag{A2}$$

Setting the reference time $t_{\text{ref}} = T/2$ (The choice of $t_{\text{ref}}$ does not affect the residual distribution in the stacked image, as it only shifts the absolute position of the accumulated signal), the compensated position of the object in the $n$-th frame is:

$$x_{\text{comp}}(t_n) = x(t_n) - v_{opt} \tau_n \tag{A3}$$

Where $\tau_n = t_n - t_{\text{ref}}$, Substituting Equations (A1) and (A2) into Equation (A3):

$$x_{\text{comp}}(t_n) = \frac{1}{2} a \tau_n^2 + C \tag{A4}$$

Where $C = x_0 + v_0 t_{\text{ref}} + \frac{1}{2} a t_{\text{ref}}^2$ is a constant representing the position at the reference time. In the stacked image, the superposed signal from all $N$ frames produces a peak at the centroid of the compensated position distribution. The degradation of the stacked peak SNR depends on how widely the individual frame signals are spread around this centroid. Therefore, we measure the residual displacement relative to the centroid $\bar{x}_{\text{comp}}$:

$$\bar{x}_{\text{comp}} = \frac{1}{N}\sum_{n=0}^{N-1} x_{\text{comp}}(t_n) = \frac{a}{2}\cdot\frac{1}{N}\sum_{n=0}^{N-1}\tau_n^2 + C \tag{A5}$$

For uniformly spaced frames, $\frac{1}{N}\sum_{n=0}^{N-1}\tau_n^2 = \Delta t^2(N^2-1)/12 \approx T^2/12$ for $N \gg 1$. Therefore:

$$\bar{x}_{\text{comp}} \approx \frac{aT^2}{24} + C \tag{A6}$$

The residual $\delta_n$, which is the deviation of the $n$-th compensated position from the centroid, is:

$$\delta_n = x_{\text{comp}}(t_n) - \bar{x}_{\text{comp}} = \frac{a}{2}\left(\tau_n^2 - \frac{T^2}{12}\right) \tag{A7}$$

The extrema of $|\delta_n|$ occur at the observation boundaries ($n=0$ and $n=N-1$, where $|\tau_n| = T/2$) and at the midpoint ($|\tau_n| \approx 0$):

$$\delta_{\text{boundary}} = \frac{a}{2}\left(\frac{T^2}{4} - \frac{T^2}{12}\right) = \frac{1}{12}aT^2 \tag{A8}$$

$$\delta_{\text{mid}} = \frac{a}{2}\left(0 - \frac{T^2}{12}\right) = -\frac{1}{24}aT^2 \tag{A9}$$

Since $\left|\frac{1}{12}aT^2\right| > \left|\frac{1}{24}aT^2\right|$, the maximum residual displacement $\delta_{\text{max}}$ occurs at the observation boundaries and equals:

$$\delta_{\text{max}} = \frac{1}{12}|a|T^2 \tag{A10}$$

**Appendix B:**

The simulation procedure is outlined as follows:
  1. Ephemeris Data Acquisition: We utilize the ephemeris data of a real cislunar space object over a specific time interval. The data consists of Right Ascension (RA) and Declination (Dec) coordinates with a temporal resolution of one second.
  2. Coordinate Transformation: The celestial coordinates are converted into image pixel coordinates $(x, y)$ by applying a gnomonic projection (tangent plane projection) followed by scaling based on the detector's pixel scale (1.67"/pixel).
  3. Image Generation: Each data point (at one-second intervals) corresponds to a single simulated image frame. The object is modeled as a point source with a

Gaussian PSF. To simulate realistic signal characteristics, photon noise following a Poisson distribution is incorporated into the object's PSF.

4. Noise Modeling: A comprehensive noise model is applied to each image, including background photon noise and dark current noise (both following a Poisson distribution), as well as readout noise (following a Gaussian distribution).

5. Shift-and-stack: The sequence of simulated images is then processed using the shift-and-stack algorithm (linear and quadratic) to evaluate its performance.

The simulation parameters, such as the pixel scale, dark current, and readout noise, are configured based on the specifications of the CMOS camera mounted on the Multi-Application Survey Telescope Array (MASTA) telescope at the Purple Mountain Observatory (PMO) (the relevant parameters of this device are listed in Table 4). These parameters were selected to ensure consistency, as the experimental validation in section 4 utilizes real observational data acquired by this specific instrument.

Table 4. Detailed information of the MASTA camera.

| Parameter | Value |
| --- | --- |
| Image resolution | 8120×8120 |
| Pixel size | 10μm |
| Plate scale | 1.67"/pixel |
| Field of view(FOV) | $3.75°×3.75°$ |
| Dark current | <0.05$e^-$ pixel$^-$ s$^-$ at -25℃ |
| Readout noise | <2$e^-$ |

**Reference**

[1] Baker-McEvilly B, Bhadauria S, Canales D, et al. A comprehensive review on Cislunar expansion and space domain awareness[J]. Progress in Aerospace Sciences, 2024, 147: 101019..

[2] Frueh C, Howell K, DeMars K J, et al. Cislunar space situational awareness[C]//31st AIAA/AAS Space Flight Mechanics Meeting. 2021, 144.

[3] Muhammad A, Wang Y, Wang H. A comprehensive review of emerging trends in lunar science exploration research pertaining to cislunar missions safety[J]. Journal of Space Safety Engineering, 2026.

[4] Chhabra A, Christensen I, Beeson R. Cislunar Environmental Outcomes: A Framework for Development and Evaluation of Long-Term Cislunar Orbital Debris Mitigation Policies[J]. Available at SSRN 6298717.

[5] Klonowski M, Owens-Fahrner N, Heidrich C, et al. Cislunar space domain awareness architecture design and analysis for cooperative agents[J]. The Journal of the Astronautical Sciences, 2024, 71(5): 47.

[6] Frueh C, Little B, McGraw J. Optical sensor model and its effects on the design of sensor networks and tracking[C]//Advanced Maui Optical and Space Surveillance Technologies Conference (AMOS). 2019..

[7] Vendl J K, Holzinger M J. Cislunar periodic orbit analysis for persistent space object detection capability[J]. Journal of Spacecraft and Rockets, 2021, 58(4): 1174-1185.

[8] Horwood J T, Aragon N D, Poore A B. Gaussian sum filters for space surveillance: theory and simulations[J]. Journal of Guidance, Control, and Dynamics, 2011, 34(6): 1839-1851.

[9] Bloom A, Wysack J, Griesbach J D, et al. Space and ground-based SDA sensor performance comparisons[C]//Advanced Maui Optical and Space Surveillance Technologies Conference. Maui Economic Development Board Maui, HW, 2022: 1-11.

[10] Zimmer P, McGraw J T, Ackermann M R. Cislunar SSA/SDA from the lunar surface: COTS imagers on commercial landers[C]//The Advanced Maui Optical and Space Surveillance Technologies (AMOS) Conference, Maui, HI. 2021: 14-17.

[11] Shao M, Nemati B, Zhai C, et al. Finding very small near-earth asteroids using synthetic tracking[J]. The Astrophysical Journal, 2014, 782(1): 1.

[12] Gladman B, Kavelaars J J, Petit J M, et al. The Structure of the Kuiper Belt: Size Distribution and RadialExtent[J]. The Astronomical Journal, 2001, 122(2): 1051.

[13] Bernstein G M, Trilling D E, Allen R L, et al. The size distribution of trans-Neptunian bodies[J]. The Astronomical Journal, 2004, 128(3): 1364.

[14] Parker A H, Kavelaars J J. Pencil-beam surveys for trans-Neptunian objects: novel methods for optimization and characterization[J]. Publications of the Astronomical Society of the Pacific, 2010, 122(891): 549.

[15] Zhai C, Shao M, Saini N S, et al. Accurate ground-based near-Earth-asteroid astrometry using synthetic tracking[J]. The Astronomical Journal, 2018, 156(2): 65.

[16] Zhai C, Shao M, Saini N, et al. Near-earth object observations using synthetic tracking[J]. Publications of the Astronomical Society of the Pacific, 2024, 136(3): 034401.

[17] Burdanov A Y, de Wit J, Brož M, et al. JWST sighting of decametre main-belt asteroids and view on meteorite sources[J]. Nature, 2025, 638(8049): 74-78.

[18] Tyson J A, Guhathakurta P, Bernstein G M, et al. Limits on the surface density of faint Kuiper Belt objects[C]//American Astronomical Society Meeting Abstracts. 1992, 181: 06.10.

[19] Cochran A L, Levison H F, Stern S A, et al. The discovery of halley-sized kuiper belt objects using hst[J]. arXiv preprint astro-ph/9509100, 1995.

[20] Gladman B, Kavelaars J J, Nicholson P D, et al. Pencil-beam surveys for faint trans-neptunian objects[J]. The Astronomical Journal, 1998, 116(4): 2042-2054.

[21] Allen R L, Bernstein G M, Malhotra R. The edge of the solar system[J]. The Astrophysical Journal, 2001, 549(2): L241.

[22] Fraser W C, Kavelaars J J. The Size Distribution of Kuiper belt objects for D ≥ 10 km[J]. The Astronomical Journal, 2009, 137(1): 72-82..

[23] Wang B, Zhao H B, Li B. Detection of faint asteroids based on image shifting and stacking method[J]. Acta Astronomica Sinica, 2017, 58(5): 49.

[24] Heinze A N, Metchev S, Trollo J. Digital tracking observations can discover asteroids 10 times fainter than conventional searches[J]. The Astronomical Journal, 2015, 150(4): 125.

[25] Zhai C, Shao M, Nemati B, et al. Detection of a faint fast-moving near-earth asteroid using the synthetic tracking technique[J]. The Astrophysical Journal, 2014, 792(1): 60.

[26] Zhai C, Ye Q, Shao M, et al. Synthetic tracking using ztf deep drilling data sets[J]. Publications of the Astronomical Society of the Pacific, 2020, 132(1012): 064502.

[27] Masci F J, Laher R R, Rusholme B, et al. The zwicky transient facility: Data processing, products, and archive[J]. Publications of the Astronomical Society of the Pacific, 2019, 131(995): 018003.

[28] Burdanov A Y, Hasler S N, de Wit J. GPU-based framework for detecting small Solar system bodies in objected exoplanet surveys[J]. Monthly Notices of the Royal Astronomical Society, 2023, 521(3): 4568-4578.

[29] Heinze A N, Metchev S. Precise Distances for Main-Belt Asteroids in Only Two Nights[J]. The Astronomical Journal, 2015, 150(4): 124.

[30] Parker A H, Kavelaars J J. Pencil-beam surveys for trans-Neptunian objects: novel methods for optimization and characterization[J]. Publications of the Astronomical Society of the Pacific, 2010, 122(891): 549.